\documentclass[aps,twocolumn,showpacs]{revtex4}
\usepackage[dvips]{graphicx}
\begin{document}
\title{A Fibonacci atomic chain with side coupled quantum dots: 
crossover from a singular continuous to a continuous spectrum and 
related issues}

\author{Arunava Chakrabarti$^1$ and Samar Chattopadhyay$^{2}$}

\affiliation{$^1$Department of Physics, University of Kalyani, Kalyani,  
West Bengal-741235, India. \\
$^2$Department of Physics, Maulana Azad College \\ 8, Rafi Ahmed Kidwai Road, 
Kolkata - 700013, India.}

\begin{abstract}
Interaction of bound states with a singular continuous 
spectrum is studied using a one dimensional Fibonacci quasicrystal 
as a prototype example.
Single level quantum dots are attached from a side to a 
subset of atomic sites of the quasiperiodic chain. 
The proximity of the dots to the chain is modeled by introducing a tunnel hopping 
between a dot and the backbone. 
It is shown that, 
depending upon the proximity of the side coupled dot, the spectrum of an 
infinite quasiperiodic chain can display radical changes from its purely 
one dimensional characteristics. Absolutely continuous parts in the spectrum 
can be generated as well as isolated resonant eigenstates whose positions in the 
spectrum are sensitive to the proximity of the quantum dots. The cycles of the matrix map 
and the two terminal transport are discussed in details.
\end{abstract}

\maketitle

\section{Introduction}
The physics of the condensed matter and materials science in the 
last couple of decades has been largely dominated by the mesoscopic 
and nano-scale systems~\cite{supriya}-\cite{ventra}. 
The advancement of lithographic techniques 
together with the use of the instruments such as the scanning 
tunnel microscope (STM) has enabled experimentalists to examine 
the physical properties of tailor made geometries and 
test their potential as future nano-electronic devices. Such studies 
have been important not only from the point of view of possible 
applications, but also due to the fact that they play a crucial role 
in understanding the effect of quantum coherence in the electronic 
transport as the size of the system is reduced below the phase coherence 
length of the electrons~\cite{supriya}.  

A large section of the existing literature in this field 
deals with the effect of 
the quantum dots (QD), single, or an array of them, 
coupled from a side, on the spectral 
and transport properties of quantum nano-wires (QW) or quantum rings (QR)
~\cite{aldea}-\cite{zhang}. The studies include the investigation of 
the effect of inter-dot coupling in a side coupled double dot system~\cite{aldea}, 
the tunability of the Fano-Kondo effect in a double QD unit~\cite{chung}, 
and a variety 
of phase coherent electronic transport studies~\cite{vasudevan}-\cite{zhang}.
The theoretical studies, in several cases, have been motivated by eperiments on 
electronic transport in QW systems that were coupled to QD's from a 
side~\cite{sasaki,kobayashi}.     

One ubiquitous phenomenon that is manifestly evident in all such 
studies on quantum transport is the occurrence of Fano effect~\cite{fano}-\cite{arunava2}. 
The Fano effect arises when a bound state 
`interacts' with a continuum, and is typically observed in the transmission 
spectrum of one dimensional QW systems when a single QD, or a cluster of them 
is attached to the QW from one side~\cite{orellana,pouthier,arunava1}. The 
transmission spectrum is marked with asymmetric lineshapes dictated 
by the formula~\cite{fano},
\begin{equation}
{\cal F}(\epsilon) = \frac{(\epsilon+\eta)^{2}}{\epsilon^{2}+1}
\end{equation}
where, $\epsilon=(E-E_R)/(\Gamma/2)$ with $E_R$ being the `resonance
energy' and $\Gamma$ the
line width. $\eta$ is the asymmetry parameter.

Inspite of the considerable volume of work existing in this field, 
a practically unaddressed issue, to the best of our knowledge, is how 
seriously does the presence of a {\it bound state} caused by the attachment 
(from one side) of a QD or an assembly of them, influence a singular 
continuous spectrum. The localized state(s) can in principle, be seated 
anywhere in the spectrum, and a singular 
continuous spectrum having a multifractal distribution of gaps may be 
severly affected by these. This is the central motivation behind the present work.

We choose to work with a Fibonacci quasiperiodic chain that is  
a classic example of a {\it one dimensional} quasicrystal presenting a 
singular continuous spectrum~\cite{shechtman}-\cite{macia2}.
Qusicrystals have been established as the third {\it ordered} phase
of the solid state~\cite{shechtman}. These are systems that are intermediate
between a perfectly periodic system, and a completely random one.
Over almost three decades, the physical properties of these strange
systems have remained under active consideration, both from the standpoints
of fundamental physics, and technological applications~\cite{kohmoto1}-\cite{macia2}.

Spectral properties of a Fibonacci quasicrystal (FQC) are exotic.
For example, its energy spectrum, in
general, is singular continuous (a Cantor set) with measure zero. 
The spectrum exhibits a variety of scaling behavior~\cite{kohmoto1}-\cite{naumis}. 
The wave functions are neither
periodic in the Bloch sense, nor are they exponentially localized as it
happens in a completely random sequence of potentials~\cite{anderson}.

Such
peculiarities in the spectral properties have prompted researchers to
propose and investigate realistic problems related to the localization of
light in quasiperiodically ordered layered 
dielectrics~\cite{sutherland1,sutherland2}, 
flux pinning~\cite{nori,kemmler}, quasiperiodic optical 
lattices~\cite{niu}-\cite{sarma}, 
plasmon excitation in aperiodically ordered dielectric layers~\cite{albu},  
and many other cases. Experiments have also been performed to test the
basic predictions of a purely one dimensional theory and to explore the
novelties of these systems~\cite{bayindir,hayashida}. 
The present day nanotechnology makes it possible  
to fabricate a lattice of QD's, and quantum wells with practically any desired  
geometry. Even stable and rigid (Carbon) atomic chains have been experimentally 
realized~\cite{jin}. In several recent communications, aperiodically ordered 
metal nano-particle arrays~\cite{negro1}-\cite{negro3}, and aperiodic arrays of 
QD's~\cite{kapulkina} have been addressed. In every case, the quasiperiodic backbone 
plays a key role in controlling the physical properties of the system.

\begin{figure}[ht]
{\centering \resizebox*{8cm}{4cm}{\includegraphics{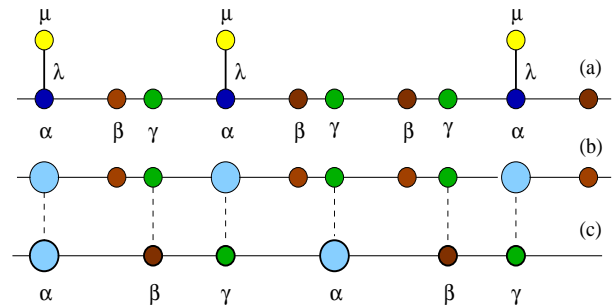}}\par}
\caption{(Color online). (a) A portion of an infinite Fibonacci chain with an adatom attached 
to every $\alpha$-site. (b) The adatom is `folded' into the backbone to generate
an `effective' $\alpha$-site, and (c) the decimation renormalization scheme.}
\label{lattice1}
\end{figure}

Such an environment has inspired us to undertake a detailed investigation 
in exploring the role of QD's side coupled 
to a Fibonacci quasiperiodic atomic chain. The dots are the  
{\it single level} QD's in the spirit of Kubala and K\"{onig}~\cite{kubala}, 
and are attached to a subset of 
atomic sites in an infinite chain (Fig.~\ref{lattice1}a).  

The results are quite extraordinary. 
Using a tight binding Hamiltonian for non-interacting, spinless electrons 
and a real space renormalization group (RSRG) scheme,  
we show that, when a 
single atomic site (equivalent to a single level QD) is attached to each of a subset of 
sites in the FQC, the energy spectrum can exhibit {\it absolutely continuous} 
sub-bands depending on the strength of the coupling of the side coupled atomic site 
to the Fibonacci backbone. The coupling represents the {\it proximity} of the adatom to the 
backbone, which can be controlled at will. There are other instances when the presence of 
adatoms can give rise to resonant (extended) eigenstates which were never present in the 
purely one dimensional model. As a result, a state that was 
{\it critical}~\cite{kohmoto1} in the purely $1$-d case, may turn out to be 
a resonant tunneling one, with a suitable proximity of the adatoms.
The matrix maps~\cite{kohmoto2}, 
typical of a Fibonacci sequence, are 
also controlled by the proximity of the adatoms. Such observations 
provide the first step towards answering the basic questions raised 
regarding the influence of bound states on a singular continuous spectrum.  
Consequently, the side coupled dots have profound effect 
on the two terminal electronic transport across a Fibonacci nanocluster. 
The system can be  
be useful in designing possible tunnel devices with a quasiperiodic backbone.     

In what follows, we describe our results. Section II contains the model and the 
principal methods of investigation. In section III we present the numerical results 
and discussion. Section IV elaborates the role of the adatom-backbone coupling 
in controlling the six cycles of the matrix map, and conclusions are drawn in section V.
\section{The Model and the method}
\subsection{The Hamiltonian}

To describe the system we use a tight-binding framework. In Wannier 
basis the Hamiltonian reads, 
\begin{equation}
H  = \sum_{i}\epsilon_{i} c_{i}^{\dagger} c_{i}
+\sum_{\langle ij \rangle} t_{ij} \left[c_{i}^{\dagger} 
c_{j} 
+ h.c. \right] 
\label{equ1}
\end{equation}
where, $\epsilon_{i}$ is the on-site energy of an electron at 
the site $i$  and $t_{ij}$ is the 
nearest-neighbor hopping strength. A binary Fibonacci chain 
comprising of two letters $L$ and $S$ is grown recursively 
following the growth rule~\cite{kohmoto1} $L \rightarrow LS$, and $S \rightarrow L$, 
beginning with $L$. The Fibonacci family in successive generations appear as, 
$G_1: L$, $G_2: LS$, $G_3: LSL$, $....$ and so on. We consider the letters $L$ and 
$S$ to represent two kinds of {\it bonds}. The on-site potential $\epsilon_i$ takes on three 
values depending on its nearest neighbor configuration, viz, it is $\epsilon_\alpha$ when 
flanked by two $L$-bonds on both sides, $\epsilon_\beta$ when it is between a $L-S$ pair, and 
$\epsilon_\gamma$ for the $S-L$ combination (see Fig.~\ref{lattice1}(a)). The nearest 
neighbor hopping is $t_L$ and $t_S$ when the electron hops across a $L$ or a $S$ bond 
respectively. The tunnel hopping, connecting a Fibonacci site and the adatom will be 
designated by $t_{ij} = \lambda$. The on-site potential of the side coupled dot is 
$\epsilon_\mu$.

\subsection{The matrix and the trace maps}

For the sake of completeness, let us remind ourselves the basics of the problem of 
determination of the eigenvalue spectrum and the eigenfunctions 
of a purely one dimensional FQC. It is efficiently 
handled by the transfer matrices $M_\ell$ corresponding to the $\ell$th generation 
Fibonacci approximant~\cite{kohmoto2}. The matrices of three consecutive generations are 
recursively coupled through the relation, 
\begin{equation}
\bf M_\ell = \bf M_{\ell-2} \bf M_{\ell-1}
\label{matrixmap}
\end{equation}
with $\bf M_1 = \bf M_\alpha$, $\bf M_2 = \bf M_\gamma \bf M_\beta$, 
and $\bf M_3 = \bf M_\alpha \bf M_\gamma \bf M_\beta$~\cite{kohmoto2}.
The transfer matrices for the individual sites read, 
\begin{eqnarray}
\bf M_\alpha & = & \left(\begin{array}{cc}
(E-\epsilon_{\alpha})/t_L & -1 \\
1 & 0 \end{array}\right) \nonumber \\
\bf M_\beta & = & \left(\begin{array}{cc}
(E-\epsilon_{\beta})/t_S & -t_L/t_S \\
1 & 0 \end{array}\right) \nonumber \\
\bf M_\gamma & = & \left(\begin{array}{cc}
(E-\epsilon_{\gamma})/t_L & -t_S/t_L \\
1 & 0 \end{array}\right)
\label{matrices}
\end{eqnarray}

The {\it allowed} eigenvalues are obtained from the condition $|x_\ell| \le 2$ as 
$\ell \rightarrow \infty$, where, 
$x_\ell = Tr \bf M_\ell$, and is obtained recursively from the equation 
\begin{equation}
x_{\ell+1} = x_\ell x_{\ell-1} - x_{\ell-2}
\label{map}
\end{equation}
with appropriate initial values of $x_\ell$ depending on the model~\cite{kohmoto2}.  
It is well known~\cite{kohmoto1}-\cite{macia2} that, as the 
generation index $\ell \rightarrow \infty$, every energy that 
one hits upon, corresponds to an escaping orbit of the trace map Eq.~\ref{map},  
and the spectrum turns into a Cantor set with a gap in the vicinity of every energy.
Such a spectrum corresponds to eigenfunctions thar are neither {\it localized} in an 
exponential way, nor are they {\it extended} in the Bloch sense. This fact is 
mathematically described by an {\it invariant} quantity given by, 
\begin{equation}
I = \frac{1}{4} ( x_\ell^2 + x_{\ell-1}^2 + x_{\ell-2}^2 - 
x_\ell x_{\ell-1} x_{\ell-2} - 4)
\label{invariant}
\end{equation}
The above invariant becomes equal to zero for a perfectly periodic 
chain of atoms, while it is infinitely large when one calculates it 
for a randomly disordered 1-d lattice. The {\it zero} of the Fibonacci 
invariant thus corresponds to extended eigenfunctions~\cite{kohmoto1}, 
a fact that is crucial for our case. 
\subsection{The RSRG scheme}

The self similarity inherent in the FQC structure makes the use of a real 
space renormalization group (RSRG) decimation scheme a natural choice to 
unravel the spectral features. 
The scheme is illustrated in Fig.~\ref{lattice1}. First, the side coupled 
QD's are `folded' in to the $\alpha$-sites to create effective $\tilde\alpha$-sites 
with potential $\epsilon_{\tilde\alpha} = \epsilon_\alpha + \lambda^2/(E-\epsilon_\mu)$.
This process is shown in Fig.~\ref{lattice1}(b).
The $\beta$-sites are then decimated to obtain a scaled version of the original chain, 
as shown in Fig.~\ref{lattice1}(c).
The decimation method relies on the use of an infinite set of 
difference equations~\cite{samar}, 
\begin{equation}
(E - \epsilon_j) \psi_j = t_{j,j+1} \psi_{j+1} + t_{j,j-1} \psi_{j-1}
\label{difference}
\end{equation} 
with $\epsilon_j = \epsilon_{\tilde\alpha}$, $\epsilon_\beta$ or $\epsilon_\gamma$ 
as appropriate, and $t_{j,j\pm 1} = t_L$ or $t_S$. 
The renormalized values of the on-site potentials and the hopping integrals are given by, 
\begin{eqnarray}
\epsilon_{\tilde\alpha}' & = & \epsilon_\gamma + \frac{t_L^2 + t_S^2}{E - \epsilon_\beta} \nonumber \\
\epsilon_\beta' & = & \epsilon_\gamma + \frac{t_S^2}{E - \epsilon_\beta} \nonumber \\
\epsilon_\gamma' & = & \epsilon_{\tilde\alpha} + \frac{t_L^2}{E - \epsilon_\beta} \nonumber \\
t_L' & = & \frac{t_L t_S}{E - \epsilon_\beta} \nonumber \\
t_S' & = & t_L
\label{recursion}
\end{eqnarray}

The above set of recursion relations may be used to obtain the local density of states (LDOS) 
at any $j$-th site through the relation $\rho_j = (-1/\pi) Im G_{jj}$ where, $G_{jj}$ is the 
local diagonal Green's function at the chosen site and, is obtained from the relation 
$G_{jj} = (E + i \eta - \epsilon_j^*)^{-1}$. $i\eta$ is the small imaginary part one needs to 
add to the energy $E$, and $\epsilon_j^*$ is the {\it fixed point} value of the 
relevant on-site potential, and is obtained from the set of Eq.~\ref{recursion}, when 
$t_L$ and $t_S$ flow to zero under RSRG iterations~\cite{southern}.
\begin{figure}[ht]
{\centering \resizebox*{12cm}{11cm}
{\includegraphics [angle=0] {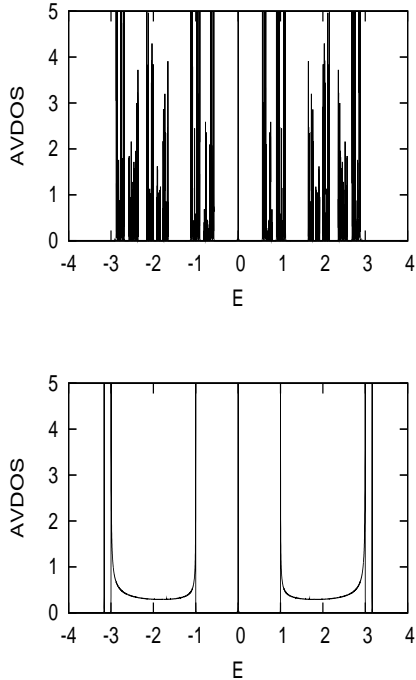}}\par}
\caption{Average density of states 
of an infinite Fibonacci quasicrystal in the transfer model, when, 
a single QD is attached to 
every $\alpha$ site. $\lambda = 1$ in the top panel and  
$\lambda = \sqrt{3}$ in the bottom panel. 
We have set $\epsilon_\alpha = 
\epsilon_\beta = \epsilon_\gamma = \epsilon_\mu = 0$, and $t_L = 1$ and $t_S = 2$.}  
\label{avdos}
\end{figure}

\subsection{The transmission coefficient}
To obtain the two terminal transmission coefficient of an $\ell$-th generation 
FQC, the sample is clamped between two semi-infinite perfectly conducting 
leads~\cite{schreiber}. The leads are modeled in the tight binding scheme by 
a uniform on-site potential $\epsilon_0$ (set equal to zero everywhere in this 
calculation), and a constant nearest neighbor hopping integral $t_0$ (set equal to 
unity throughout).
The transmission coefficient of the $\ell$-th generation FQC is 
then obtained from the formula~\cite{schreiber},
\begin{equation}
T = \frac{4 - E^2}{(E z_\ell/2 - y_\ell)^2 + x_\ell^2 (1 - E^2/4)}
\label{transmission}
\end{equation}
where, $x_\ell = \bf M_\ell(1,1) + \bf M_\ell(2,2)$, 
$y_\ell = \bf M_\ell(2,1) - \bf M_\ell(1,2)$,
and $z_\ell = \bf M_\ell(1,1) - \bf M_\ell(2,2)$. The values of $x_\ell$, $y_\ell$ and 
$z_\ell$ at any $\ell$-th generation are obtained from their respective recursion relations, viz, 
$x_{\ell} = x_{\ell-1} x_{\ell-2} - x_{\ell-3}$, $y_{\ell} = x_{\ell-1} y_{\ell-2} + y_{\ell-3}$ 
and $z_{\ell} = x_{\ell-1} z_{\ell-2} + z_{\ell-3}$ with appropriate intial 
values~\cite{schreiber}.
\section{Results and discussions}
\subsection{Absoloutely continuous subbands}

Let us refer to 
Fig.~\ref{lattice1} (a). A single adatom marked $\mu$ and 
with on-site potential $\epsilon_\mu$ is attached 
to every $\alpha$-site of the chain. The FQC-adatom coupling is $\lambda$. The 
effect of the adatom is easily taken care of by defining a {\it renormalized} 
potential at the $\alpha$-site (Fig.~\ref{lattice1}b). 
That is, the intial value of the potential at 
the $\alpha$-site is not just $\epsilon_\alpha$, but, 
$\tilde \epsilon_\alpha = \epsilon_\alpha + \lambda^2/(E - \epsilon_\mu)$.  
With this modification, we work out the invariant (given by 
Eq.~\ref{invariant}) of the trace map in Eq.~\ref{map}
for a purely transfer model~\cite{kohmoto2} with $\epsilon_\alpha = \epsilon_\beta =
\epsilon_\gamma$, $t_L = \tau$, and $t_S = R \tau$. 
If we choose $\epsilon_\mu = \epsilon_\alpha$, the invariant 
remains independent of energy, and reads, 
\begin{equation}
I = \frac{[\lambda^2 - (R^2 - 1) \tau^2]^2}{4 R^2 \tau^4}
\label{invariant2}
\end{equation}
This immediately leads to a situation where one can have a {\it zero} of the 
invariant independent of the electron energy $E$. 
The zero of the invariant in case of a FQC should correspond to extended eigenstates
~\cite{kohmoto1}.
Setting $I = 0$ we obtain 
$\lambda = \pm \tau \sqrt{R^2 - 1}$. This gives us a measure of the {\it proximity} 
of the adatom for which one should get extended eigenstates in a FQC irrespective 
of energy. 

Do these extended eigenstates form a {\it band} ? To answer this question, we 
work out the commutator 
$[\bf M_{\gamma\beta},\bf M_{\tilde\alpha}]$, where, 
$\bf M_{\gamma\beta} = \bf M_\gamma \bf M_\beta$, and 
$\bf M_{\tilde\alpha}$ are the transfer matrices for the $\gamma\beta$ cluster and the 
`renormalized' $\alpha$ (now called $\tilde\alpha$) atoms respectively. 
The result is, 
\begin{equation}
[\bf M_{\gamma\beta},\bf M_{\tilde\alpha}] = 0
\label{comm}
\end{equation}
for $\lambda = \pm \tau \sqrt{R^2 - 1}$, independent of the electron energy $E$.
It is to be appreciated that, this value of $\lambda$ is the same as obtained 
by forcing the inavariant to vanish.
The 
vanishing of the commutator implies that, for the above choice of the tunnel hopping 
$\lambda$, the $\gamma\beta$ cluster and the $\tilde\alpha$ atoms can even be 
arranged in a periodically alternating manner, representing an ordered binary alloy. 
The energy band in this case consists of continuous distribution of 
eigenvalues with a gap separating the continuous sub-clusters.
As a result of the commutation~Eq.\ref{comm}, the electron will `feel' 
no essential difference of the FQC with an ordered binary alloy, and the 
energy spectra of the original FQC should therefore be 
identical to that of a periodic arrangement of the constituent clusters 
$\tilde\alpha$ and $\gamma\beta$ placed in an alternating fashion.
{\it Continuous subbands of extended states} in the spectrum,  
when $\lambda = \pm \tau \sqrt{R^2 - 1}$, is therefore an obvious result for the 
stubbed FQC.
\begin{figure}[ht]
{\centering \resizebox*{12cm}{11cm}
{\includegraphics [angle=0] {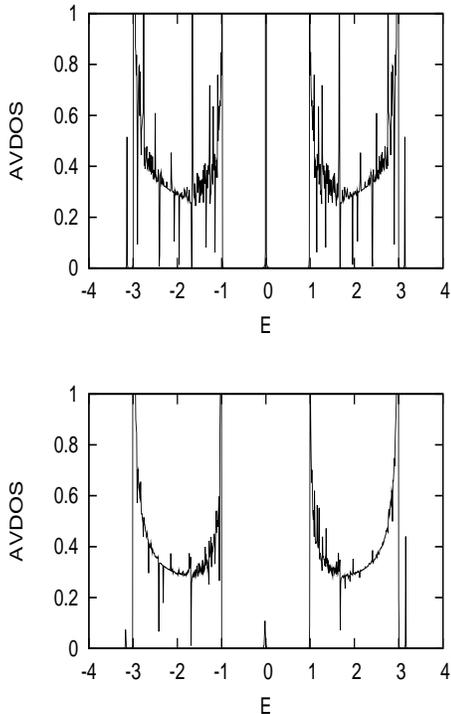}}\par}
\caption{Average density of states of a 
Fibonacci quasicrystal in the transfer model.  
A single QD is attached to 
every $\alpha$ site. We have chosen ($\lambda = 1.69736$, $\epsilon_\mu = 0$)
(top panel), 
and ($\lambda = \sqrt{3}$, $\epsilon_\mu = -0.02$) (bottom panel)
representing a $2\%$ deviation from the exact values (with respect to Fig.~\ref{avdos} 
(bottom panel) of any 
one of the parameters $\lambda$ 
and $\epsilon_\mu$ respectively. Once again we have 
set $\epsilon_\alpha = \epsilon_\beta = \epsilon_\gamma = 0$, 
$t_L = 1$ and $t_S = 2$.} 
\label{fig3}
\end{figure}

To confirm this we present 
in Fig.~\ref{avdos} 
the average density of states (AVDOS) of a FQC with adatoms attached to 
every $\alpha$-site.. 
We have set $\epsilon_\alpha = \epsilon_\beta = \epsilon_\gamma = 0$,  
$t_L = 1$ and $t_S = 2$. along the Fibonacci backbone. 
The site potential of the adatom is taken as $\epsilon_\mu = 0$, and 
the value of the FQC-adatom tunnel hopping $\lambda$ has been 
set equal to $1$ and $\sqrt{3}$ respectively in the two figures.  
For $\lambda = 1$ 
there appears a notable change in the spectrum 
compared to the usual three-subband spectrum of a 
standard FQC~\cite{kohmoto2}. The sub-bands get shifted and the 
eigenvalues cluster in a different shape. 
There is a central peak at $E = 0$ that corresponds to a sharply localized 
eigenstate, and is a consequence of the attachment of the adatoms.
But otherwise, the spectrum retains the typical fragmented, self similar 
character of a 1-d FQC.

The remarkable part of the figure is 
the case with $\lambda = \sqrt{3}$ which corresponds to $I = 0$. The spectrum is 
{\it absolutely continuous} within two sub-bands.  
The central localized state remains pinned at $E = 0$ however.
Two additional localized states show up immediately beyond the sub-bands 
on either side.
The overall spectral character has been cross checked  
by explicitly working out the trace map. The completely gapless character 
of the spectrum in the range $-3 < E < -1$ and $1 < E < 3$ 
suggests that one should have all the eigenstates extended 
in these energy regimes when  
$\lambda = \sqrt{3}$.  

The recursion relations Eq.~\ref{recursion} reconfirm the extendedness of an 
eigenstate corresponding to an eigenvalue falling within the continuum  
in the spectrum. The hopping integrals $t_L$ and $t_S$ remain non-zero under 
successive RSRG iteration for an indefinite number of loops if the chosen 
energy is picked up from within the continuum. This implies that, at any length scale 
there is a non-zero connectivity betweeen the neighboring sites at that scale, and 
hence the corresponding state is of extended character~\cite{southern}.

\begin{figure}[ht]
{\centering \resizebox*{15.5cm}{12.5cm}
{\includegraphics [angle=0] {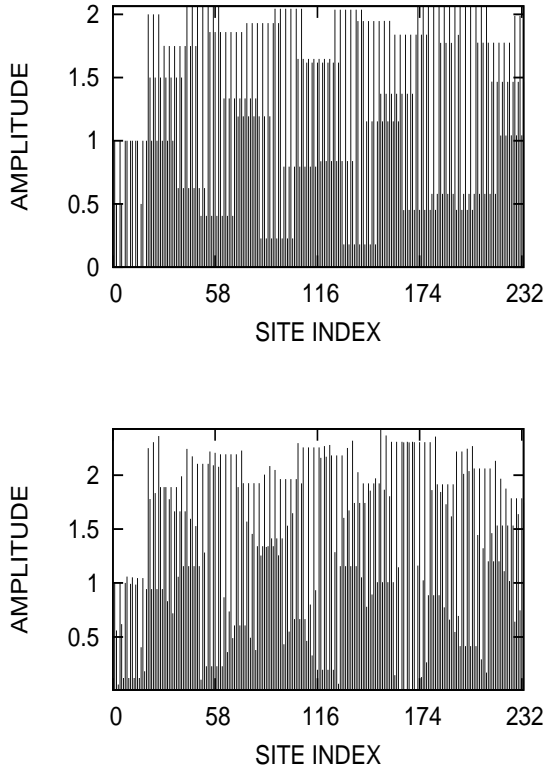}}\par}
\caption{$|\psi_n|$ plotted for $232$ sites
of a Fibonacci chain with adatom attached to every $\alpha$-site.  
We have chosen $\lambda = \sqrt{3}$ (top) and  
$\lambda = 1.69736$ (bottom) showing that the extended nature of the 
wave function persists even with a $2\%$ variation in $\lambda$. 
$\epsilon_\mu = 0$ in both the figures and $E = 2$, selected in the 
{\it continuous band} of the spectrum. Other parameters are the same as in 
Fig.2.}
\label{figampli}
\end{figure}

Before we end this subsection, it's important to appreciate that, 
the origin of the continuous bands of extended states is the 
commutation of the matrices $\bf M_{\gamma\beta}$, and $\bf M_\alpha$.
Therefore, any quasiperiodic, or even 
disordered geometric arrangement of $\alpha$ (coupled to a single adatom), 
and the $\beta\gamma$ 
pair, under suitable conditions such as the 
above will lead to an identical energy spectrum and hence, extended eigenstates.
We have tested this with other kinds of aperiodic chains which under RSRG 
give rise to various kinds of recursion relations between the Hamiltonian parameters.
In every case, the bands with $\lambda = \sqrt{3}$ is the same. In this respect, 
such microscopically different aperiodic geometries can be brought under one 
{\it universality class}, using a suitable side coupled array of QD's. The transport 
behavior of course, is sensitive to the recursion relations, and its structural details 
will be different. 
\subsection{Stability of the continuous spectrum}

A major concern 
from the standpoint of an experimentalist will be the stability of the 
continuum against a possible deviation of $\lambda$ from its 
exact numerical value as obtained from Eq.~\ref{invariant2}, or 
a variance in the value of $\epsilon_\mu$. This is related to 
any error in fixing up the exact proximity of the adatom to the $\alpha$-sites, 
or an error in controlling the potential of the attached QD by a gate voltage. 
To this end, we have extensively studied the AVDOS spectrum by varying $\lambda$ 
from its value of $\sqrt{3}$. For large deviations from $\lambda = \sqrt{3}$, 
the spectrum gets back to the familiar three sub-band Cantor set like structure 
typical to a FQC. However, for small deviations the continua persist, though with 
a ruggedness in the landscape, which becomes more and more dominant with 
increasing value of $\lambda$. 

To illustrate, we have shown in Fig.~\ref{fig3} 
the AVDOS  for $\lambda = 1.69736$, $\epsilon_\mu = 0$ (top)  
, and for $\lambda = \sqrt{3}$, $\epsilon_\mu = -0.02$ (bottom) for 
the entire range of the spectrum. 
These values 
are deviations by $2\%$  
from the {\it critical} values, viz, $\lambda = \sqrt{3}$, or $\epsilon_\mu = 0$. 
The smooth AVDOS observed in Fig.\ref{avdos} (bottom) 
gets distorted by numerous oscillations in Fig.~\ref{fig3}. 
But, the continuous distribuion of eigenvalues still persists over smaller 
energy intervals. The patches of continua are also observed to survive even 
when the deviation is as large as $6\%$ from either $\lambda = \sqrt{3}$, or 
$\epsilon_\mu = 0$, or even when both the parameters vary within a reasonable 
uncertainty in their values.
This stability is also reflected in the 
amplitude distribution along the FQC backbone (Fig.~\ref{figampli}).
The energy is chosen to be $E = 2$, that is, from within a continuous subband.
Oscillating but non-decaying profile of the amplitudes for arbitrarily large 
system size is observed. We have shown (top panel) the result for $232$ sites, 
with $\epsilon_\alpha = \epsilon_\beta = \epsilon_\gamma = \epsilon_\mu = 0$, 
$t_L =1$, $t_S = 2$, and for $\lambda = \sqrt{3}$. 
In the bottom panel we present the case where  
the FQC-adatom hopping $\lambda$ deviates by $2\%$ from its {\it critical} 
value of $\lambda = \sqrt{3}$. The extendedness prevails over extremely 
large system size, though we again show the results for just $232$ sites here.

Thus, even within the natural 
experimental uncertainty, it should be possible to 
observe a crossover in the transport characteristics from the poorly 
conducting to a metallic 
one in a Fibonacci array with side coupled quantum dots, by 
controlling the proximity, or the potential of the dots with respect 
to the backbone.

\subsection{The transmission coefficient}
\begin{figure}[ht]
{\centering \resizebox*{15.0cm}{13.5cm}
{\includegraphics [angle=0] {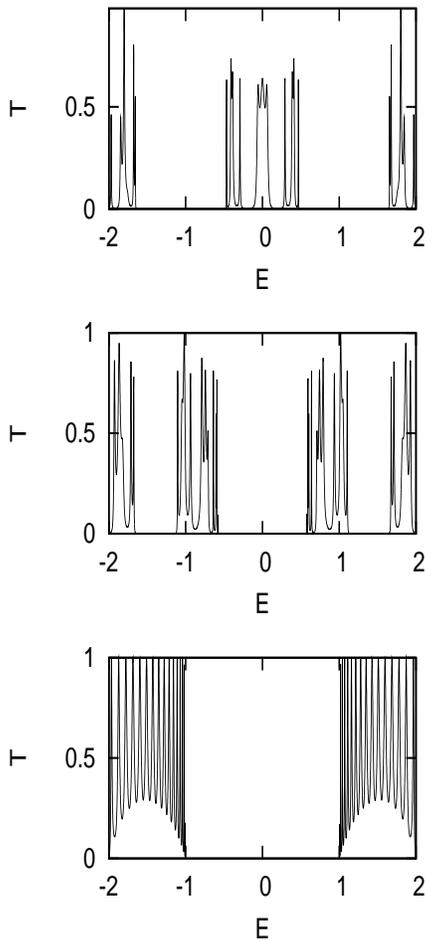}}\par}
\caption{Transmission spectrum of a $9$th generation 
Fibonacci quasicrystal ($55$ bonds) with 
(a) no side attachement, (b) 
an adatom attached to every $\alpha$-site in 
the bare length scale, and $\lambda = 1$, and (c) same as (b), but now with 
$\lambda = \sqrt{3}$.}
\label{transport}
\end{figure}

The obervations made in the density of states spectrum are corroborated 
by the corresponding calculation of the transmission coefficient 
of finite but arbitrarily large systems using Eq.~\ref{transmission}. 
In Fig.~\ref{transport} we 
present the transmission spectra for $\lambda = 0$ (Fig.~\ref{transport}(a), no side atoms), 
$\lambda = 1$ (Fig.~\ref{transport}(b)) and $\lambda = \sqrt{3}$ (Fig.~\ref{transport}(c)) 
for the transfer model with $\epsilon_i = \epsilon_\mu = 0$, $t_L = 1$, and $t_S = 2$.
The three sub-band clustering of the pure FQC in (a) evolves into a 
four sub-band form. $\lambda = \sqrt{3}$ clearly shrinks the 
entire spectrum into two continuous zones of high transmission coefficients. 
The sharply localized state at $E = 0$ of course, never contributes. 
This  
reflects what we have already discussed in the context of the density of states.
\section{Six cycles of the matrix map and the role of $\lambda$}

Extended eigenstates can also be traced which are 
related to a six cyclic behavior of the matrix map
viz, $\bf M_{\ell+6} = \bf M_\ell$, $\ell \ge 1$~\cite{kohmoto1}-
\cite{kohmoto2}. In a 
previous work~\cite{samar} it has been shown that a six cycle of 
the matrix map given by Eq.~\ref{matrixmap} is caused by resonances 
occuring in the clusters $(\gamma\beta - \gamma\beta$, $\alpha-\alpha)$ in 
the bare length scale, or through the resonances taking place in pairs like 
$(\alpha\gamma\beta - \alpha\gamma\beta$, $\gamma\beta - \gamma\beta)$ 
in one step renormalized lattice. Bigger clusters are easily identified 
from higher order renormalized version of the FQC. 

As already mentioned, in the six cyclic cases, 
$\bf M_\ell = \bf M_{\ell+6}$, for one or more special values of energy. The value of 
$\ell$ determines the length scale at which the six cycle of the map 
begins to show up, and depends on the construction of the resonating 
clusters~\cite{samar}. 
The energy eigenvalue corresponding to the six cycle of the matrix map
is extracted as a common root of the equations, 
$Tr \bf M_{\gamma\beta}(n) = 0$, and 
$Tr \bf M_\alpha(n) = 0$.
$n$ denotes the stage of renormalization, and it 
provides the length scale at which the resonating clusters 
are identified. 

In the FQC with attached QD's, the six cycle energies can be obtained almost at will 
by fixing the adatom at suitable places in the original lattice. For 
example, in the bare length scale $Tr \bf M_{\gamma\beta} = 0$ gives 
\begin{equation}
E = \frac{1}{2} \left [ \epsilon_\gamma + \epsilon_\beta \pm 
\sqrt{(\epsilon_\gamma - \epsilon_\beta)^2 + 
4 (t_S^2 + t_L^2)} \right ]
\label{first}
\end{equation}
while, the solution of $Tr \bf M_{\tilde\alpha} = 0$ gives, 
\begin{equation}
E = \frac{1}{2} \left [ (\epsilon_\alpha + \epsilon_\mu) 
\pm \sqrt{(\epsilon_\alpha - \epsilon_\mu)^2 + 
4 \lambda^2} \right ]
\label{second}
\end{equation}
One immediately finds that, for a given set of $(\epsilon_\alpha, 
\epsilon_\beta, \epsilon_\gamma, t_L, t_S)$, and for a certain value 
of $\epsilon_\mu$, one can tune the tunnel hopping $\lambda$ so as 
to satisfy Eqs.~\ref{first} and ~\ref{second} simultaneously. 
By choosing $\epsilon_\alpha = \epsilon_\beta = \epsilon_\gamma = \epsilon_\mu$, 
Eq.~\ref{first} provides $E = \epsilon_\alpha \pm \sqrt{t_L^2 + t_S^2}$
which is independent of $\lambda$. This energy value can be 
made equal to that obtained from the second equation, viz, 
Eq.~\ref{second}
by choosing $\lambda = \pm \sqrt{t_S^2 + t_L^2}$. This gives 
an estimate of the {\it proximity} of an adatom to the $\alpha$-sites 
in the original lattice that will be leading to six cycles of the matrix 
maps at special values of energies. We have extracted numerous such energy values 
from various scales of length. Each such energy corresponds to an extended eigenstate 
of the system in the sense that, the hopping integrals remain non-zero for 
an indefinite number of RSRG iterations.
\section{Conclusions}

In conclusion, we have examined the consequence of an interaction 
between bound states and a singular continuous energy spectrum by 
fixing isolated single level quantum dots as adatoms on 
specific lattice points of an infinite quasi-periodic 
Fibonacci array of atomic sites. The energy spectrum of system exhibits a remarkable 
transformation from a completely fragmented Cantor set nature with  
measure zero to one with continuous distribution of eigenvalues, similar 
to that of an ordered binary alloy, by judiciously choosing the 
QD-FQC tunnel hopping. 
One thus achieves an {\it almost insulating} to a metallic behavior of the 
system when the Fermi energy is located at suitable parts of the spectrum.
The spectral crossover is found to be robust against a possible 
variation in the values of the tunnel hopping integral, or the 
potential of the adatoms.
Other 
established properties of a FQC are also inspected and are found to be 
sensitive to the values of the tunnel hopping which gives us an 
estimate of the proximity of the adatoms to the lattice. The two terminal 
transmission coefficient is evaluated using the standard recursive 
algorithm, and has been shown to corroborate our findings regarding the 
energy spectrom of sun a stubbed Fibonacci array of quantum dots. 
Detailed analysis of the system at different scales of length can be 
made utilizing the real space renormalization group methods. This is 
under investigation, and the results will be reported in due course.
\vskip 0.3in
\noindent
                                                                            
\end{document}